\documentclass[prb,preprint,showpacs,preprintnumbers,amsmath,amssymb,aps]{revtex4-1}
\usepackage{graphicx}
\usepackage{dcolumn}
\usepackage{bm}
\usepackage{color}
\def\Im {\mbox{Im}}
\def\be{\begin{equation}}       \def\ee{\end{equation}}
\def\bea{\begin{eqnarray}}      \def\eea{\end{eqnarray}}
\def\bes{\begin{subequations}}  \def\ees{\end{subequations}}

\def\dag{\dagger}
\def\non{\nonumber}

\def\k{_{\bf k}}
\def\km{_{-\bf k}}

\begin{document}

\title{Spin dynamics of antiferromagnetically coupled bilayers -- the case of Cr$_2$TeO$_6$}
\author{Kingshuk Majumdar}
\email{majumdak@gvsu.edu}
\affiliation{Department of Physics, Grand Valley State University, Allendale, 
Michigan 49401, USA}
\author{Subhendra D. Mahanti}
\email{mahanti@pa.msu.edu}
\affiliation{Department of Physics and Astronomy, Michigan State University, East Lansing, 
Michigan 48824, USA}

\date{\today}

\begin{abstract}
\label{abstract}
Understanding the dynamics of interacting quantum spins has been one of the active areas of condensed matter physics research.
Recently, extensive inelastic neutron scattering measurements have been carried out in an interesting class of systems,
Cr$_2$(Te, W, Mo)O$_6$. These systems consist 
of bilayers of Cr$^{3+}$ spins ($S=3/2$) with strong antiferromagnetic  inter-bilayer coupling ($J$) and tuneable intra-bilayer coupling 
($j$) from ferro (for W and Mo) to antiferro (for Te). In the limit when $J>|j|$, the system reduces to weakly interacting quantum
spin-3/2 dimers. In this paper we discuss the low-temperature magnetic 
properties of Cr$_2$TeO$_6$ systems where both intra-layer and inter-layer exchange couplings are antiferromagnetic, i.e. $J,j>0$. 
Using linear spin-wave theory we obtain the magnon dispersion, sublattice magnetization, two-magnon density of states, and 
longitudinal spin-spin correlation function.
\end{abstract}$
$

\maketitle

\section{\label{sec:Intro}Introduction}
Quantum spin fluctuations (QSF) play an important role in the low temperature properties of 
quantum antiferromagnets (QAF), particularly in systems with low spin and 
low dimension.~\cite{anderson,harris71,diep,mila} For example, the ordered moment or sublattice magnetization ($M_s$) in a 
nearest-neighbor (NN)
D-dimensional spin-1/2 Heisenberg quantum antiferromagnet (HQAF) is zero for $D=1$ (no long range order), 
0.3067$\mu_B$ for 
$D=2$,~\cite{anderson,majumdar10} and 
0.423$\mu_B$ for $D=3$ (isotropic couplings)~\cite{majumdar11a}, the latter two values obtained in a leading order approximation 
(to be discussed later in the paper). 
Thus QSF decrease with increasing $S$ and increasing $D$. The interplay of QSF and covalency induced reduction 
of $M_s$ in QAFs has also been a subject of great interest in the past, particularly in the parent compound 
of high $T_c$ superconductors (La$_2$CuO$_4$ where the Cu$^{2+}$ ions have $S=1/2$, form a square lattice 
and interact with NN isotropic Heisenberg antiferromagnetic interaction).~\cite{mahanti91}

Recently Zhu et al~\cite{zhu14} have studied the magnetic structure of an interesting class of layered magnetic 
systems containing Cr$_2$XO$_6;$ X=Te, Mo, W where the Cr$^{3+}$ (spin-3/2) magnetic ions are arranged in bilayers. 
The inter-bilayer coupling is strongly antiferromagnetic due to the presence of Cr$^{3+}-$ Cr$^{3+}$ dimers. The intra-bilayer 
couplings however can be either antiferromagnetic (AF) or ferromagnetic (F) depending 
on whether the system contains Te or W (also Mo) atom.~\cite{zhu15}  In fact the average intra-bilayer exchange can be tuned 
from one limit to the other in Cr$_2$W$_{1-x}$Te$_x$O$_6$ by changing $x$ from 0 to 1. Neutron powder 
diffraction (NPD) measurements have determined the ground state spin structure and the values of the sublattice 
magnetization. The Te compound consists of antiferromagnetic bilayers which are coupled antiferromagnetically (AF-AF). 
In contrast the W and Mo analogs consist of ferromagnetic bilayers coupled antiferromagnetically (F-AF). In addition 
to the magnetic ordering the NPD measurements also give the values of sublattice magnetization $M_s$. The values of $M_s$ 
are reduced from their 
atomic spin value $3.0\mu_B$ for Cr$^{3+}\;(M_s=g\mu_B S)$ assuming $g=2$ and quenched orbital angular momentum. 
This reduction can be due to covalency where 
the Cr $d$ orbitals hybridize 
with O $p$ orbitals and due to QSF.~\cite{mahanti91} Electronic structure calculations reveal
information about the reduction due to covalency whereas QSF-caused reduction can be calculated using a quantum Heisenberg 
spin Hamiltonian, which is one of the issues we address in this paper.

{\it Ab initio} electronic structure calculations using density 
functional theory (GGA and GGA+U)~\cite{dudarev,anisimov91,anisimov93} correctly reproduced the magnetic ordering in these three 
compounds.~\cite{zhu14} Since the orbital 
angular momentum is quenched for the Cr$^{3+}$ configuration (three electrons in the t$_{2g}$ orbitals; $S=3/2$) the magnetic 
moment comes from the spin. The estimated exchange parameters (inter-bilayer or inter-dimer exchange $J$ and 
intra-bilayer NN exchange $j$) were reasonable in view of the limitations of GGA or GGA+U approximations.~\cite{zhu14} However 
the calculated values of the sublattice magnetization ($\sim2.8\mu_B$) was very close to the ionic value ($3.0\mu_B$) indicating 
a small ($\sim 6.5\%$) covalent 
reduction of the ordered moment.~\cite{zhu14} In contrast, the experimental values are reduced to $\sim2.3\mu_B$.~\cite{zhu14} 
One possible reason for this reduction is QSF.~\cite{anderson} Such dramatic reduction in ordered moment has been seen 
in many quasi-two dimensional QAFs, a classic example being La$_2$CuO$_4$ which consists of antiferromagnetic 2D square 
lattice of $S=1/2$, where QSF reduce the sublattice magnetization ($M_s$) by $\sim 40\%$.~\cite{anderson} In the present systems the 
reduction should be smaller (at least by a factor of three) due to $S=3/2$.

\begin{figure}[httb]
\centering
(a) \includegraphics[width=2.3in,clip]{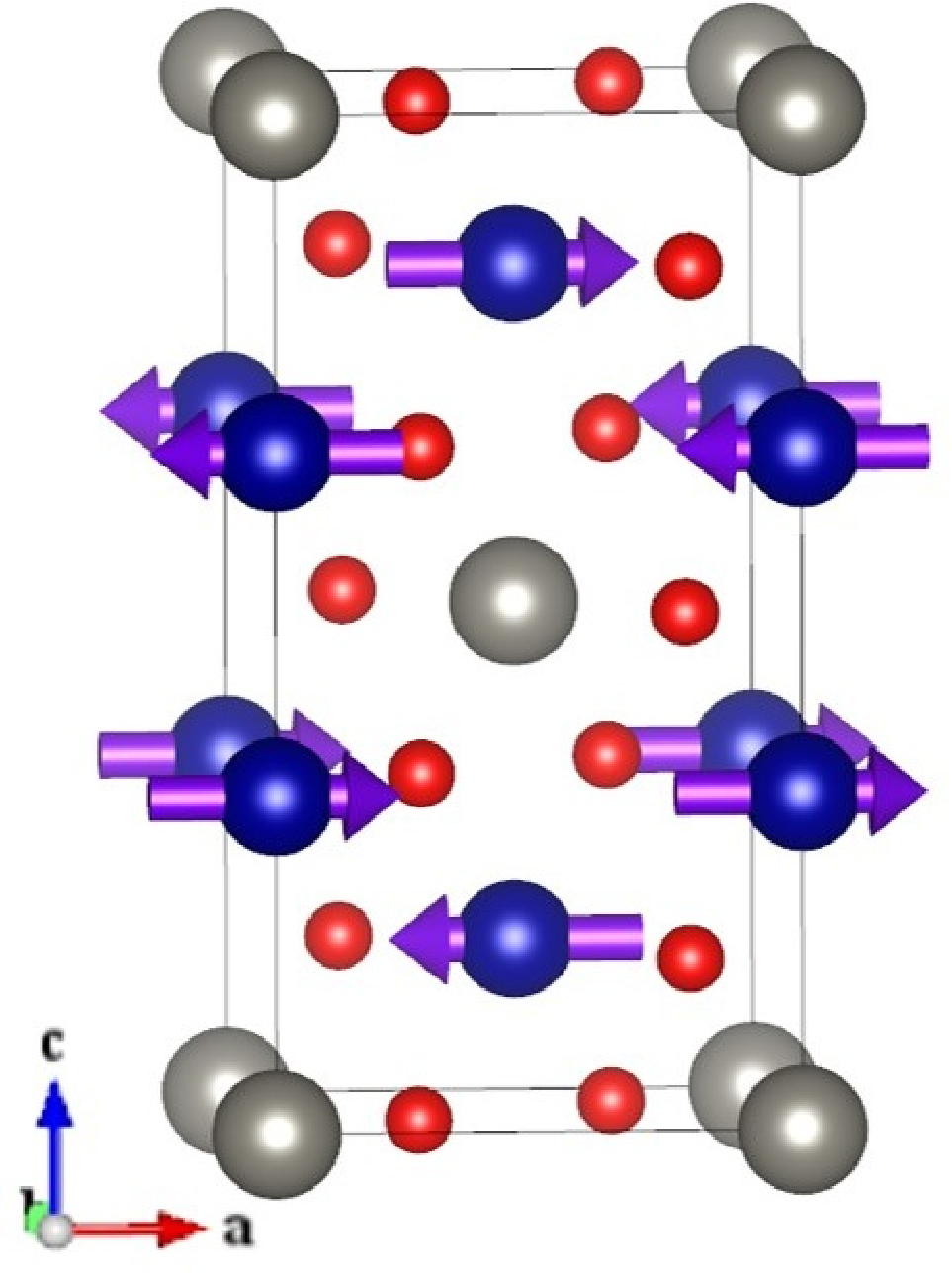}
\qquad 
(b) \includegraphics[width=3.0in,clip]{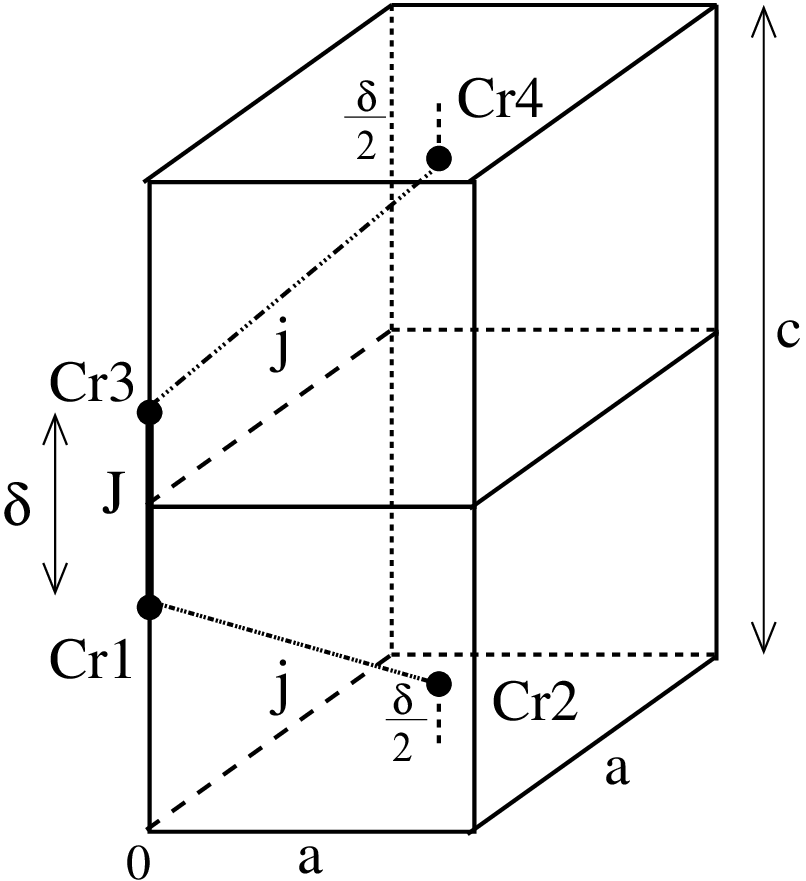}
\caption{\label{fig:CrMWstruc1} (Color online)(a) Schematic of the bilayer crystal structure of Cr$_2$TeO$_6$.
Each Cr$^{3+}$ (blue balls) bilayer is separated by a Te layer (silver balls).~\cite{zhu14,zhu15} (b) 
Positions of four chromium spins in the tetragonal unit cell of dimensions
$(a,a,c)$ are shown. The coordinates of the spins are:
Cr1: $(0,0,c/2-\delta/2)$, Cr2: $(a/2,a/2,\delta/2)$, Cr3: $(0,0,c/2+\delta/2)$, and Cr4: $(a/2,a/2,c-\delta/2)$.
Within each bilayer Cr1 (Cr3) spin is coupled with Cr2 (Cr4) spin (shown by dashed lines 
with coupling strength $j$). On the other hand, for 
the inter-bilayer NN coupling
Cr1 and Cr3 are coupled within the same tetragonal cell (shown by a thick solid line with coupling strength $J$) whereas Cr2 is coupled 
to Cr4  in the unit cell below and 
Cr4 is coupled with Cr2 in the unit cell above.}
\end{figure}

To visualize the magnetic ordering and exchange coupling in these systems we will consider the ground state spin ordering in 
Cr$_2$TeO$_6$ [see Fig.~\ref{fig:CrMWstruc1}].~\cite{kunn68} One has two bilayers (perpendicular to the $z$-axis) in the 
tetragonal unit cell $(a,a,c)$ and four Cr 
spins/unit cell. The experimental unit cell parameters are $a=4.545$\AA\; and $c=8.995$\AA\; for Cr$_2$TeO$_6$ and 
$a=4.583$\AA\; and $c=8.853$\AA \;for Cr$_2$WO$_6$.~\cite{zhu14} The distance between the inter-bilayer (NN) Cr atoms i.e. Cr1 and 
Cr3 or Cr2 and Cr4 is $\delta \sim 3.00$\AA\;$\approx c/3$, whereas the distance between 
intra-bilayer NN Cr atoms (Cr1 and Cr2 or Cr3 and Cr4) is $\sim 3.80$\AA. One bilayer contains Cr1 and Cr2 spins and the other 
contains Cr3 and Cr4 spins. The 
inter-bilayer coupling comes through Cr1-Cr3 and Cr2-Cr4 dimers, it is antiferromagnetic and its strength 
is denoted by $J$. The NN intra-bilayer coupling is between Cr3-Cr4 and Cr1-Cr2 and its strength is denoted 
by $j$. The magnitude of $j$ is considerably smaller than that of $J$ in these systems which can therefore regarded as weakly 
interacting quantum dimers. In  
Cr$_2$TeO$_6$ the intra-bilayer coupling is antiferromagnetic. In thermodynamic measurements the 
high temperature properties (for example peak in heat capacity) are determined primarily by the dimers i.e. the energy scale is 
set by the intra-dimer coupling strength $J$ whereas for the low temperature properties below the antiferromagnetic transition 
temperature 
the intra-dimer coupling $j$ is responsible for the long range order. Also it plays an important role in magnon dispersion 
and quantum spin fluctuations. Experimental values of the couplings estimated from high temperature susceptibility 
measurements are: $|J|=2.9$ meV and $|j|=0.4$ meV for Cr$_2$TeO$_6$ and 
$|J|=3.8$ meV and $|j|=0.12$ meV for Cr$_2$WO$_6$.~\cite{drillon79}

The present systems are somewhat peculiar. The inter-bilayer coupling $J$ is AF and strong. 
The intra-bilayer coupling $j$ is small and can be either F or AF. If we put $j=0$, then the system consists of non-interacting 
quantum spin spin-3/2 dimers and the ground state consists of a product of dimer singlet states and there is no long range 
order (LRO). If on the other hand, $J=0$, the system consists of non-interacting bilayers. When $j$ is AF, the system is 
similar to the cuprates~\cite{mahanti91} described above but the QSF reduces $M_s$ by only 13\% because $S=3/2$. When $j$ is F, there 
is no 
QSF reduction of $M_s$. An important question is how $J$ and $j$ interfere with each other. To study this interesting 
problem we have calculated the magnetic excitations and QSF for the interacting quantum spin dimer problem using a  
Heisenberg quantum Hamiltonian.

In the current article we discuss the low temperature magnetic properties and the role of QSF in
Cr$_2$TeO$_6$ systems using a three parameter $(J,j,j^\prime)$ spin-3/2 quantum Heisenberg Hamiltonian. Here $j^\prime$ is the 
next nearest neighbor (NNN) intra-bilayer exchange coupling which is weakly ferromagnetic. Specifically, we obtain the 
magnon energy dispersion, sublattice magnetization, longitudinal spin-spin correlation function, and its powder-averaged intensity. In a forthcoming 
paper we will discuss the spin dynamics of Cr$_2$WO$_6$ and Cr$_2$MoO$_6$ systems, where the intra-bilayer coupling is ferromagnetic and inter-bilayer coupling is
antiferromagnetic. The paper is organized as follows: we define the spin Hamiltonian in Section~\ref{sec:model} and discuss the formalism 
using linear spin-wave theory~\cite{majumdar10, majumdar12} for the AF-AF bilayers  
(Section~\ref{sec:model}). In Section~\ref{sec:results} we present the results for the magnon dispersion, sublattice 
magnetization, two-magnon density of states, longitudinal spin-spin correlation function, and its powder-average. 
We conclude our paper with a brief conclusion in Section~\ref{sec:conclusions}.

\section{\label{sec:model} Formalism}

It is well-known that quantum fluctuations play a significant role in the magnetic properties~\cite{anderson,harris71} and phase 
diagram of the system at zero
temperature.~\cite{diep,mila,chubukov92,majumdar13,majumdar12,majumdar11b,majumdar11a,majumdar10} Here we investigate the role of 
quantum fluctuations on the stability of the N\'{e}el
state by calculating the magnon spectrum and see how these excitations reduce the sublattice magnetization from its N\'{e}el state 
value. There are several anaytical and numerical methods to study the low temperature properties of quantum magnets.~\cite{diep}
One such method is the spin-wave theory (SWT), which has proved to be a very effective to the study of 
quantum magnets described by the Heisenberg hamiltonian especially for dimensions $D\ge 2$ and large spin $S$.~\cite{anderson} SWT provides accurate results for many physical
quantities even for the difficult case of spin-1/2 quantum Heisenberg antiferromagnet in two dimensions. Our quantum bilayer spin system is 
three dimensional and the Cr$^{3+}$ ions have spin-3/2 -- thus SWT method is well suited for the present bilayer spin system. 
In the SWT formalism, we first express the fluctuations around the classical long range ordered antiferromagnetic ground state in 
terms of the bosonic operators using the Holstein-Primakoff (HP) representation.~\cite{HP} The quadratic term in magnon
operators corresponds to the linear spin-wave theory (LSWT), whereas the higher-order terms e.g. quartic terms (Ref. ~\onlinecite{majumdar10})
represent interactions between magnons which we ignore in our current work. 

\subsection{\label{MD} Magnon Dispersion}
As discussed in Section~\ref{sec:model} there are four chromium atoms in the tetragonal unit cell as shown in 
Fig.~\ref{fig:CrMWstruc1}. Within each bilayer the 
spins of Cr1 and Cr3 are coupled antiferromagnetically with the 
spins of Cr2 and Cr4 respectively through the intra-bilayer NN coupling. On the other hand,
the NN Cr1 and Cr3 spins belonging to different bilayers are coupled antiferromagnetically within the same unit cell 
whereas Cr2 spin is coupled antiferromagnetically to Cr4 spin in the unit cell below and Cr4 spin is coupled with Cr2 in the unit cell 
above. 

The Heisenberg Hamiltonian of the system with antiferromagnetic intra- and inter-bilayer couplings
$j$ and $J$ ($j,\;J>0$) has the form
\bea
{\cal H}_{\rm NN} &=& j\sum_{n=1}^{N_z} \sum_{\langle i,j\rangle} \Big[ {\bf S}_{in}^{(1)A}\cdot {\bf S}_{jn}^{(2)B}
      +{\bf S}_{in}^{(3)B}\cdot {\bf S}_{jn}^{(4)A}\Big]\non \\
&+&J\sum_{n=1}^{N_z} \sum_{i} \Big[ {\bf S}_{in}^{(1)A}\cdot {\bf S}_{in}^{(3)B}
      +\frac 1{2}\{ {\bf S}_{in}^{(2)B}\cdot {\bf S}_{in-1}^{(4)A}+{\bf S}_{in}^{(4)A}\cdot {\bf S}_{in+1}^{(2)B}\}\Big],
\label{ham-partial}
\eea
where $n$ represents the $n$-th unit cell index along the $z$-direction ($N_z$ is the number of cells along $z$-direction) 
and $i,j$ are nearest-neighbor (NN) sites within the same 
bilayer ($\langle i, j\rangle$ implies that each bond is counted once). For example,
${\bf S}_{in}^{(1)A}$ represents the spin of Cr1 at site $i$ in the $n$-th unit cell whereas ${\bf S}_{jn+1}^{(2)B}$ 
represents spin of Cr2 in the $(n+1)$-th unit cell. A and B are indices for sublattices A (spin-up) and B (spin-down).
Within the bilayer spins of Cr1, Cr4 are in A sublattice and  Cr2, Cr3 are in B sublattice. The factor of 1/2 in Eq.~\ref{ham-partial}
takes into account double counting of the exchange coupling between Cr2 and Cr4 while doing the sum over 
$n$ (along the $z$-direction). In addition to the NN intra-layer 
and inter-layer couplings we also add next-to-nearest neighbor (NNN) ferromagnetic coupling $j^\prime$. The full Hamiltonian is then:
\be
{\cal H}={\cal H}_{\rm NN}+{\cal H}_{\rm NNN},
\label{ham-AF}
\ee
with
\be
{\cal H}_{\rm NNN}=-j^\prime \sum_{n=1}^{N_z}\sum_{\langle \langle i,j\rangle \rangle}\Big[{\bf S}_{in}^{(1)A}\cdot {\bf S}_{jn}^{(1)A}
+{\bf S}_{in}^{(4)A}\cdot {\bf S}_{jn}^{(4)A}+{\bf S}_{in}^{(2)B}\cdot {\bf S}_{jn}^{(2)B}+{\bf S}_{in}^{(3)B}\cdot {\bf S}_{jn}^{(3)B}\Big].
\ee
Above $\langle \langle i,j \rangle \rangle$ refer to NNN interaction (bond) between the spins, each bond is counted once, and $j^\prime >0$
(ferromagnetic interaction). 
This spin Hamiltonian is mapped onto an equivalent Hamiltonian
of interacting bosons by expressing the spin operators in terms of bosonic creation 
and annihilation operators 
$a^\dag, a$ for ``up'' sites on sublattice A (and $b^\dag, b$ for ``down'' sites on  sublattice 
B)  using the  Holstein-Primakoff representation~\cite{HP}
\begin{eqnarray}\label{holstein}
S_{in}^{+ A} &\approx& \sqrt{2S}a_{in},\;\;\;
S_{in}^{- A} \approx \sqrt{2S}a_{in}^\dag,\;\;\;
S_{in}^{z A}= S-a^\dag_{in}a_{in}, \non \\ 
S_{jn}^{+ B} &\approx& \sqrt{2S}b_{jn}^\dag,\;\;\;
S_{jn}^{- B} \approx \sqrt{2S}b_{jn}, \;\;\;
S_{jn}^{z B} = -S+b^\dag_{jn}b_{jn}. 
\end{eqnarray}
Substituting Eq.~\eqref{holstein} into Eq.~\eqref{ham-AF} 
we expand the Hamiltonian perturbatively in powers of $1/S$ up to the quadratic term as
\be
{\cal H} = {\cal H}_{\rm cl}+{\cal H}_0 + \cdots,
\ee 
where,
\begin{subequations}
 \label{HAF} 
\begin{eqnarray}
{\cal H}_{\rm cl} &=& -2jNS^2\Big[ 4(1+\eta^\prime)+\eta \Big], \label{HAF1}  \\
{\cal H}_{0} &=& jS\sum_{n=1}^{N_z} \sum_{\langle i,j\rangle} \Big[a_{in}^{(1)\dag}a_{in}^{(1)}+a_{in}^{(4)\dag}a_{in}^{(4)} 
+b_{jn}^{(2)\dag}b_{jn}^{(2)}+b_{jn}^{(3)\dag}b_{jn}^{(3)} \non \\
&+& a_{in}^{(1)}b_{jn}^{(2)}+a_{in}^{(4)}b_{jn}^{(3)}
+a_{in}^{(1)\dag}b_{jn}^{(2)\dag}+a_{in}^{(4)\dag}b_{jn}^{(3)\dag}\Big] \non \\
&+& JS\sum_{n=1}^{N_z} \sum_{i} \Big[a_{in}^{(1)\dag}a_{in}^{(1)}+b_{in}^{(3)\dag}b_{in}^{(3)} 
+ a_{in}^{(1)}b_{in}^{(3)}+a_{in}^{(1)\dag}b_{in}^{(3)\dag} \non \\
&+& \frac 1{2}\Big\{ a_{in-1}^{(4)\dag}a_{in-1}^{(4)}+b_{in}^{(2)\dag}b_{in}^{(2)}
+a_{in}^{(4)\dag}a_{in}^{(4)}+b_{in+1}^{(2)\dag}b_{in+1}^{(2)}  \non \\
&+& a_{in}^{(4)}b_{in+1}^{(2)}+a_{in}^{(4)\dag}b_{in+1}^{(2)\dag} 
+a_{in-1}^{(4)}b_{in}^{(2)}+a_{in-1}^{(4)\dag}b_{in}^{(2)\dag}\Big\}\Big]\non \\
&+&j^\prime S\sum_{n=1}^{N_z} \sum_{\langle\langle i,j\rangle\rangle} \sum_{p=1,4}\Big[a_{in}^{(p)\dag}a_{in}^{(p)}+a_{jn}^{(p)\dag}a_{jn}^{(p)} 
-a_{in}^{(p)\dag}a_{jn}^{(p)}-a_{in}^{(p)}a_{jn}^{(p)\dag}\Big] \non \\
&+& j^\prime S\sum_{n=1}^{N_z} \sum_{\langle\langle i,j\rangle\rangle} \sum_{p=2,3}\Big[b_{in}^{(p)\dag}b_{in}^{(p)}+b_{jn}^{(p)\dag}b_{jn}^{(p)} 
-b_{in}^{(p)\dag}b_{jn}^{(p)}-b_{in}^{(p)}b_{jn}^{(p)\dag}\Big].\label{HAF2} 
\end{eqnarray}
\end{subequations}
${\cal H}_{\rm cl}$ above is just a number 
representing the classical ground state (mean-field) energy - so we do not discuss it further as
it is not relevant for the quantum fluctuations. $H_0$ in Eq.~\eqref{HAF2} is the quadratic part of the Hamiltonian.
In Eq.~\eqref{HAF1}, the parameters $\eta=J/j$, $\eta^\prime=j^\prime/j$ and $N=N_xN_yN_z$ is the total number of unit cells. Next the 
real space Hamiltonian 
is transformed to momentum space using the Fourier transformation (FT) for each $\ell$-th spin:
\be
a_{in}^{(\ell)}=\frac 1{\sqrt{N}}\sum_{{\bf k}}e^{i{\bf k \cdot R_{in}^{(\ell)}}}a_{\bf k}^{(\ell)},\;\;\;
b_{in}^{(\ell)}=\frac 1{\sqrt{N}}\sum_{{\bf k}}e^{-i{\bf k \cdot R_{in}^{(\ell)}}}b_{-\bf k}^{(\ell)}.
\ee
Furthermore we have rescaled the operators $a,\;b$ as 
\begin{subequations}
\bea
a\k^{(1)} &\equiv& e^{-ik_z \delta/2} a\k^{(1)},\;\;
a\k^{(4)}\equiv e^{-ik_z \delta/2} a\k^{(4)}, \\
b\km^{(2)}&\equiv& e^{-ik_z \delta/2} b\km^{(2)},\;\;
b\km^{(3)}\equiv e^{-ik_z \delta/2} b\km^{(3)}, 
\eea
\end{subequations}
where $\delta$ is the intra-dimer separation [Fig.~\ref{fig:CrMWstruc1}].
In momentum space the quadratic Hamiltonian becomes:
\begin{eqnarray}
{\cal H}_0 &=& jS(4+\eta)\sum\k \kappa \k\Big[ \Big(a\k^{{(1)}\dag} a\k^{(1)}+a\k^{{(4)}\dag} a\k^{(4)}+b\km^{{(2)}\dag} b\km^{(2)}
+b\km^{{(3)}\dag} b\km^{(3)}\Big) \nonumber \\
&+& \gamma_{1 \bf k}\Big(a^{(1)}\k b^{(2)}\km+a^{(4)}\k b^{(3)}\km \Big)+ 
\gamma_{1 \bf k}^*\Big(a^{(1)\dag}\k b^{(2)\dag}\km+a^{(4)\dag}\k b^{(3)\dag}\km \Big) \nonumber \\
&+& \gamma_{2 \bf k}\Big(a^{(1)}\k b^{(3)}\km+a^{(4)}\k b^{(2)}\km \Big)+ 
\gamma_{2 \bf k}^*\Big(a^{(1)\dag}\k b^{(3)\dag}\km+a^{(4)\dag}\k b^{(2)\dag}\km \Big)
\Big],
\end{eqnarray}
with,
\begin{subequations}
 \bea
 \gamma_{1 \bf k}&=& \frac 4{4+\eta} \frac {e^{ik_z c/2}\cos (k_xa/2)\cos (k_y a/2)}{1+\gamma_{3\bf k}},  \\
 \gamma_{2 \bf k} &=& \frac {\eta}{4+\eta}\frac 1{1+\gamma_{3\bf k}}, \\
 \gamma_{3 \bf k}&=& \frac {4\eta^\prime}{4+\eta}\Big[1-\frac 1{2}(\cos(k_x a)+\cos (k_y a))\Big], \\
 \kappa\k &=& 1+\gamma_{3\bf k}.
\eea
\end{subequations}
Finally, we diagonalize the quadratic part $H_0$ by transforming the 
operators $a_{\bf k}$ and $b_{\bf k}$ to magnon operators 
$\alpha_{\bf k}$ and $\beta_{\bf k}$ using the generalized Bogoliubov (BG)~\cite{bogoliubov, colpa} transformations:
\begin{subequations}
\begin{eqnarray}
 a\k^{(1)} &=& \frac 1{\sqrt 2}\Big[C_1 \alpha \k^{(1)}-S_1\beta\km^{(1)\dag}+ C_2 \alpha \k^{(2)}-S_2\beta\km^{(2)\dag}\Big],  \\
 b\km^{(2)} &=& \frac 1{\sqrt 2}\Big[\zeta_1^*(-S_1 \alpha \k^{(1)\dag}+C_1\beta\km^{(1)})+ 
 \zeta_2^*(-S_2 \alpha \k^{(2)\dag}+C_2\beta\km^{(2)})\Big], \\
 a\k^{(4)} &=& \frac 1{\sqrt 2}\Big[-C_1 \alpha \k^{(1)}+S_1\beta\km^{(1)\dag}+ C_2 \alpha \k^{(2)}-S_2\beta\km^{(2)\dag}\Big],  \\
 b\km^{(3)} &=& \frac 1{\sqrt 2}\Big[\zeta_1^*(S_1 \alpha \k^{(1)\dag}-C_1\beta\km^{(1)})+ 
 \zeta_2^*(-S_2 \alpha \k^{(2)\dag}+C_2\beta\km^{(2)})\Big],
 \end{eqnarray}
 \end{subequations}
where $C_1=\cosh(\theta_{k1}), S_1=\sinh(\theta_{k1}), C_2=\cosh (\theta_{k2}), S_2=\sinh (\theta_{k2})$, and $\zeta_{1k}, \zeta_{2k}$ 
are phase factors to be determined later. The diagonalization conditions after BG transformations are:
\be
\frac {2C_1S_1}{C_1^2+S_1^2}=\tanh (2\theta_{k1})=\vert \gamma^{-}\k \vert,\;\;\;
\frac {2C_2S_2}{C_2^2+S_2^2}=\tanh (2\theta_{k2})=\vert \gamma^{+}\k \vert,
\ee
where, $\gamma^{\pm}\k=\gamma_{1\bf k} \pm \gamma_{2 \bf k}$. We also determine the 
functions $\zeta_1$ and $\zeta_2$, which are  $\zeta_{1\bf k}= {\gamma\k^{-}}/{|\gamma\k^-|}$ and 
$\zeta_{2\bf k}={\gamma\k^{+}}/{|\gamma\k^+|}$.
The diagonalized quadratic Hamiltonian becomes:
\bea
\label{diagH}
{\cal H}_0 &=&jS(4+\eta)\sum\k\kappa\k \Big\{\omega^{(1)}_{\bf k}\Big[\alpha\k^{(1)\dag}\alpha\k^{(1)}+ \beta\km^{(1)\dag}\beta\km^{(1)}\Big] 
+ \omega^{(2)}_{\bf k}\Big[\alpha\k^{(2)\dag}\alpha\k^{(2)}+ \beta\km^{(2)\dag}\beta\km^{(2)}\Big]\Big\}\non \\
&-&jS(4+\eta)\sum\k \kappa\k\Big[\omega^{(1)}_{\bf k}+\omega^{(2)}_{\bf k}-2\Big],
\eea
where $\omega^{(1)}_{\bf k}=\Big[1-|\gamma\k^{+}|^2\Big]^{1/2}$ and $\omega^{(2)}_{\bf k}=\Big[1-|\gamma\k^{-}|^2\Big]^{1/2}$.
The second constant term in Eq.~\eqref{diagH} is the quantum-zero point energy, which contributes to the classical ground state
energy. The quasiparticle energy $E_{\bf k}^{(1,2)}$ for both $\alpha$ and $\beta$ magnon branches are given by: 
\be
E_{\bf k}^{(1,2)}=jS(4+\eta)\kappa\k \omega\k^{(1,2)}. \label{dispersion}
\ee

\subsection{\label{SM} Sublattice Magnetization}
The normalized sublattice magnetization, $m_s=M_s/M_0$ (where $M_0=g\mu_B$) for the A-sublattice can be expressed as 
\be m_s=S-\delta S, 
\label{MagAF}
\ee
where,
\be 
\delta S=\frac 1{N}\sum\k \langle a^{(1) \dag}\k a^{(1)}\k \rangle
=-\frac 1{2}+\frac 1{2N}\sum\k \frac 1{2}\Big[\frac {1}{\omega^{(1)}\k}+ \frac {1}{\omega^{(2)}\k}\Big].
\ee
$\delta S$ corresponds to the reduction of magnetization due to quantum fluctuations within LSWT and the summation over ${\bf k}$ goes over the entire
Brillouin zone corresponding to the tetragonal unit cell $(a,a,c)$.

\subsection{\label{LSSF}Longitudinal spin-spin correlation function (LSSF)}
In this section we derive the expressions for the longitudinal spin-spin correlation function (LSSF).~\cite{canali93} 
It is defined as
\be
{\cal L}_s({\bf k}, t)=\langle S_z ({\bf k}, t) S_z(-{\bf k}, 0) \rangle,
\label{lss}
\ee 
where
\be
S_z ({\bf k})=\frac 1{\sqrt {4N}}\sum_{i\mu} S_z^{i\mu} e^{-i{\bf k} \cdot ({{\bf R}_i}+\tau_\mu)}.
\ee
Here ${\bf R}_i$ is the position vector of the $i$-th unit cell and ${\bf \tau}_\mu$ are the positions of the four Cr-atoms in the unit cell. 
The position of the Cr-atoms are respectively: 
Cr1: $\tau_1=(0,0, c/2-\delta/2)$, Cr2: $\tau_2=(a/2,a/2,\delta/2)$,
Cr3: $\tau_3=(0,0, c/2+\delta/2)$, and Cr4: $\tau_2=(a/2,a/2,c-\delta/2)$ [See Fig.~\ref{fig:CrMWstruc1}]. Based on the experimental data $c=2a$ and $\delta=c/3$.
The quantity measured in neutron-scattering experiment is the Fourier transform of the time-dependent spin-correlation function
${\cal L}_s({\bf k},t)$,
\be
{\cal L}_s({\bf k}, \omega)=\int_{-\infty}^{\infty} \frac {dt}{2\pi} {\cal L}_s({\bf k}, t)e^{-i\omega t}.
\ee
The spins for each of the sublattices $1, 2, 3,4$ are defined in terms of the operators $a$ and $b$ as:
\begin{subequations}
\bea
S_{zn}^{(1)} &=& S-a_{in}^{(1)\dagger}a_{in}^{(1)},\\
S_{zn}^{(4)} &=& S-a_{in}^{(4)\dagger}a_{in}^{(4)}, \\
S_{zn}^{(2)} &=& -S+b_{jn}^{(2)\dagger}b_{jn}^{(2)},\\
S_{zn}^{(3)} &=& -S+b_{jn}^{(3)\dagger}b_{jn}^{(3)},
\eea
\end{subequations}
which after FT become:
\begin{subequations}
\bea
S_z^{(1)}({\bf k}) &=& \sqrt{4N}S\delta({\bf k}=0)-\frac 1{\sqrt {4N}}\sum_{{\bf p,q}}\delta ({\bf k}+{\bf p}-{\bf q})f_{1{\bf k}} 
a_{\bf p}^{(1)\dagger}a_{\bf q}^{(1)}, \\
S_z^{(4)}({\bf k}) &=& \sqrt{4N}S\delta({\bf k}=0)-\frac 1{\sqrt {4N}}\sum_{{\bf p,q}}\delta ({\bf k}+{\bf p}-{\bf q})f_{4{\bf k}} 
a_{\bf p}^{(4)\dagger}a_{\bf q}^{(4)}, \\
S_z^{(2)}({\bf k}) &=& -\sqrt{4N}S\delta({\bf k}=0)+\frac 1{\sqrt {4N}}\sum_{{\bf p,q}}\delta ({\bf k}+{\bf p}-{\bf q}) 
f_{2{\bf k}}b_{-{\bf q}}^{(2)\dagger}b_{-\bf p}^{(2)},\\
S_z^{(3)}({\bf k}) &=& -\sqrt{4N}S\delta({\bf k}=0)+\frac 1{\sqrt {4N}}\sum_{{\bf p,q}}\delta ({\bf k}+{\bf p}-{\bf q}) 
f_{3{\bf k}}b_{-{\bf q}}^{(3)\dagger}b_{-\bf p}^{(3)}, 
\eea
\end{subequations}
where $f_{\mu {\bf k}}=e^{-i{\bf k \cdot \tau_\mu}}$ takes into account the relative phases of the different magnetic atoms inside 
the unit cell. The total spin can now be written as:
\be
S_z({\bf k})=-\frac 1{\sqrt{4N}}\sum_{{\bf p,q}}\delta ({\bf k}+{\bf p}-{\bf q}) 
\Big\{[ f_{1\bf k}a_{\bf p}^{(1)\dagger}a_{\bf q}^{(1)}+f_{4\bf k}a_{\bf p}^{(4)\dagger}a_{\bf q}^{(4)}]
-[f_{2\bf k}b_{-{\bf q}}^{(2)\dagger}b_{-\bf p}^{(2)}+f_{3\bf k}b_{-{\bf q}}^{(3)\dagger}b_{-\bf p}^{(3)}] \Big\}.
\ee
Using BG transformations we express $S_z({\bf k})$ in terms of the magnon operators $\alpha$ and $\beta$. The result is shown in the 
Appendix~\ref{SScorr}.

There are $16 \times 16$ time-ordered Green's functions (GFs) that arise from Eq.~\eqref{lss}, of which only four 
contribute to LSSF. These four are defined as: 
\begin{subequations}
\bea
\Pi_1(t) &=& -i\langle T \beta^{(1)}_{-\bf p}(t)\alpha_{\bf q}^{(1)}(t) \alpha_{\bf p^\prime}^{(1)\dagger}(0)
\beta_{-\bf {q^\prime}}^{(1)\dagger}(0) \rangle, \\
\Pi_2(t) &=& -i\langle T \beta^{(2)}_{-\bf p}(t)\alpha_{\bf q}^{(2)}(t) \alpha_{\bf p^\prime}^{(2)\dagger}(0)
\beta_{-\bf {q^\prime}}^{(2)\dagger}(0) \rangle, \\
\Pi_3(t) &=& -i\langle T \beta^{(2)}_{-\bf p}(t)\alpha_{\bf q}^{(1)}(t) \alpha_{\bf p^\prime}^{(1)\dagger}(0)
\beta_{-\bf {q^\prime}}^{(2)\dagger}(0) \rangle, \\
\Pi_4(t) &=& -i\langle T \beta^{(1)}_{-\bf p}(t)\alpha_{\bf q}^{(2)}(t) \alpha_{\bf p^\prime}^{(2)\dagger}(0)
\beta_{-\bf {q^\prime}}^{(1)\dagger}(0) \rangle. 
\eea
\end{subequations}
These GFs can be calculated easily in the leading order which does not include magnon-magnon interactions. The imaginary parts of these
GFs are:
\begin{subequations}
\bea
-\frac 1{\pi}\;{\Im}\; \Pi_1^{(0)} (\omega) &=& \delta_{\bf {pq^\prime}}\delta_{{\bf p^\prime q}}
\delta (\omega-\omega^{(1)}_{\bf p}-\omega^{(1)}_{\bf q}), \\
-\frac 1{\pi}\;{\Im}\; \Pi_2^{(0)} (\omega) &=& \delta_{\bf {pq^\prime}}\delta_{{\bf p^\prime q}}
\delta (\omega-\omega^{(2)}_{\bf p}-\omega^{(2)}_{\bf q}), \\
-\frac 1{\pi}\;{\Im}\; \Pi_3^{(0)} (\omega) &=& \delta_{\bf {pq^\prime}}\delta_{{\bf p^\prime q}}
\delta (\omega-\omega^{(2)}_{\bf p}-\omega^{(1)}_{\bf q}), \\
-\frac 1{\pi}\;{\Im}\; \Pi_4^{(0)} (\omega) &=& \delta_{\bf {pq^\prime}}\delta_{{\bf p^\prime q}}
\delta (\omega-\omega^{(1)}_{\bf p}-\omega^{(2)}_{\bf q}).
\eea
\end{subequations}
The imaginary parts correspond to the spectral densities of the correlation function ${\cal L}_s ({\bf k}, \omega)$, which is:
\bea
{\cal L}_s({\bf k}, \omega) &=& \frac 1{4N}\Big[\sum_{\bf p} \delta (\omega-\omega^{(1)}_{\bf p}-\omega^{(1)}_{\bf p+k})
\vert {\cal D}^{11}_{\bf k, k+p} \vert^2 
+\sum_{\bf p} \delta (\omega-\omega^{(2)}_{\bf p}-\omega^{(2)}_{\bf p+k})
\vert  {\cal D}^{22}_{\bf k, k+p} \vert^2 \non \\
&+&\sum_{\bf p} \delta (\omega-\omega^{(2)}_{\bf p}-\omega^{(1)}_{\bf p+k})
\vert {\cal D}^{21}_{\bf k, k+p} \vert^2 
+\sum_{\bf p} \delta (\omega-\omega^{(1)}_{\bf p}-\omega^{(2)}_{\bf p+k})
\vert  {\cal D}^{12}_{\bf k, k+p} \vert^2\Big],  \label{spin-corr}
\eea
where the form factors $D^{ij}_{{\bf k, k+p}}$ are defined as,
\begin{subequations}
\bea
{\cal D}^{11}_{\bf k, k+p}&=& \frac 1{2}\Big\{[f_{1{\bf k}}+f_{4{\bf k}}]C_{1\bf k+p}S_{1\bf p}
-[f_{2{\bf k}}+f_{3{\bf k}}]\zeta_{1\bf p}^*\zeta_{1\bf k+p}C_{1\bf p} S_{1\bf k+p}\Big\},  \\
{\cal D}^{22}_{\bf k, k+p}&=& \frac 1{2}\Big\{[f_{1{\bf k}}+f_{4{\bf k}}]C_{2\bf k+p}S_{2\bf p}
-[f_{2{\bf k}}+f_{3{\bf k}}]\zeta_{2\bf p}^*\zeta_{2\bf k+p}C_{2\bf p} S_{2\bf k+p}\Big\}, \\
{\cal D}^{21}_{\bf k, k+p}&=& \frac 1{2}\Big\{[f_{1{\bf k}}+f_{4{\bf k}}]C_{1\bf k+p}S_{2\bf p}
+[f_{2{\bf k}}+f_{3{\bf k}}]\zeta_{2\bf p}^*\zeta_{1\bf k+p}C_{2\bf p} S_{1\bf k+p}\Big\}, \\
{\cal D}^{12}_{\bf k, k+p}&=& \frac 1{2}\Big\{[f_{1{\bf k}}+f_{4{\bf k}}]C_{2\bf k+p}S_{1\bf p}
+[f_{2{\bf k}}+f_{3{\bf k}}]\zeta_{1\bf p}^*\zeta_{2\bf k+p}C_{1\bf p} S_{2\bf k+p}\Big\}.  
\eea
\end{subequations}
Usually neutron scattering studies are done in powder samples. 
The maximum value $Q_{max}$ of the wave-vector is $Q_{\rm max}=\sqrt{(\pi/a)^2+(\pi/a)^2+(\pi/c)^2}=1.5\pi/a$, where $c=2a$. Then 
$k_xa=Qa\sin\theta\cos\phi, k_ya=Qa\sin \theta \sin \phi, k_zc=Qc \cos \theta$. 
The powder average of the longitudinal spin-spin correlation function is obtained by averaging over the angles $\theta$ and $\phi$
for a given value of $Q$:
\be
\langle {\cal L}_s (Q,\omega) \rangle = \frac 1{4\pi} \int_0^{2\pi} d\phi \int_0^{\pi} d\theta \sin \theta \;{\cal L}_s({\bf k}, \omega).
\label{powavg}
\ee
\subsection{\label{TMDOS} Two-Magnon Density of States (DOS)}
The two-magnon density of states (DOS) for the four GFs are given as:
\begin{subequations}
\bea
{\rm DOS}_{11}({\bf k}, \omega) &=& \sum_{\bf {p}}\delta (\omega-\omega^{(1)}_{\bf p}-\omega^{(1)}_{\bf k+p}), \label{dos1}\\
{\rm DOS}_{22}({\bf k}, \omega) &=& \sum_{\bf {p}}\delta (\omega-\omega^{(2)}_{\bf p}-\omega^{(2)}_{\bf k+p}), \label{dos2}\\
{\rm DOS}_{21}({\bf k}, \omega) &=& \sum_{\bf {p}}\delta (\omega-\omega^{(2)}_{\bf p}-\omega^{(1)}_{\bf k+p}), \label{dos3}\\
{\rm DOS}_{12}({\bf k}, \omega) &=& \sum_{\bf {p}}\delta (\omega-\omega^{(1)}_{\bf p}-\omega^{(2)}_{\bf k+p}).\label{dos4}
\eea
\label{dos}
\end{subequations}
For $\eta=0$, the $\alpha$ and $\beta$ branches are four-fold degenerate. In that case, 
${\rm DOS}_{11}({\bf k}, \omega)={\rm DOS}_{22}({\bf k}, \omega)
={\rm DOS}_{12}({\bf k}, \omega)={\rm DOS}_{21}({\bf k}, \omega)$. For $\eta \ne 0$ there are also symmetries between the four DOS's.
For example, with $k_x=\pi/a, k_y=k_z=0$ or $k_x=k_y=\pi/a, k_z=0$ or $k_x=k_y=0, k_z=\pi/c$, or $k_x a=k_y a=k_z c=\pi$,
DOS$_{11}=$DOS$_{12}$ and DOS$_{22}=$DOS$_{21}$. 

\section{\label{sec:results}Results: Magnon Dispersion, Sublattice Magnetization, and Longitudinal Spin-Spin correlation function}
\subsection{Magnon Energy Dispersion}
\begin{figure}[httb]
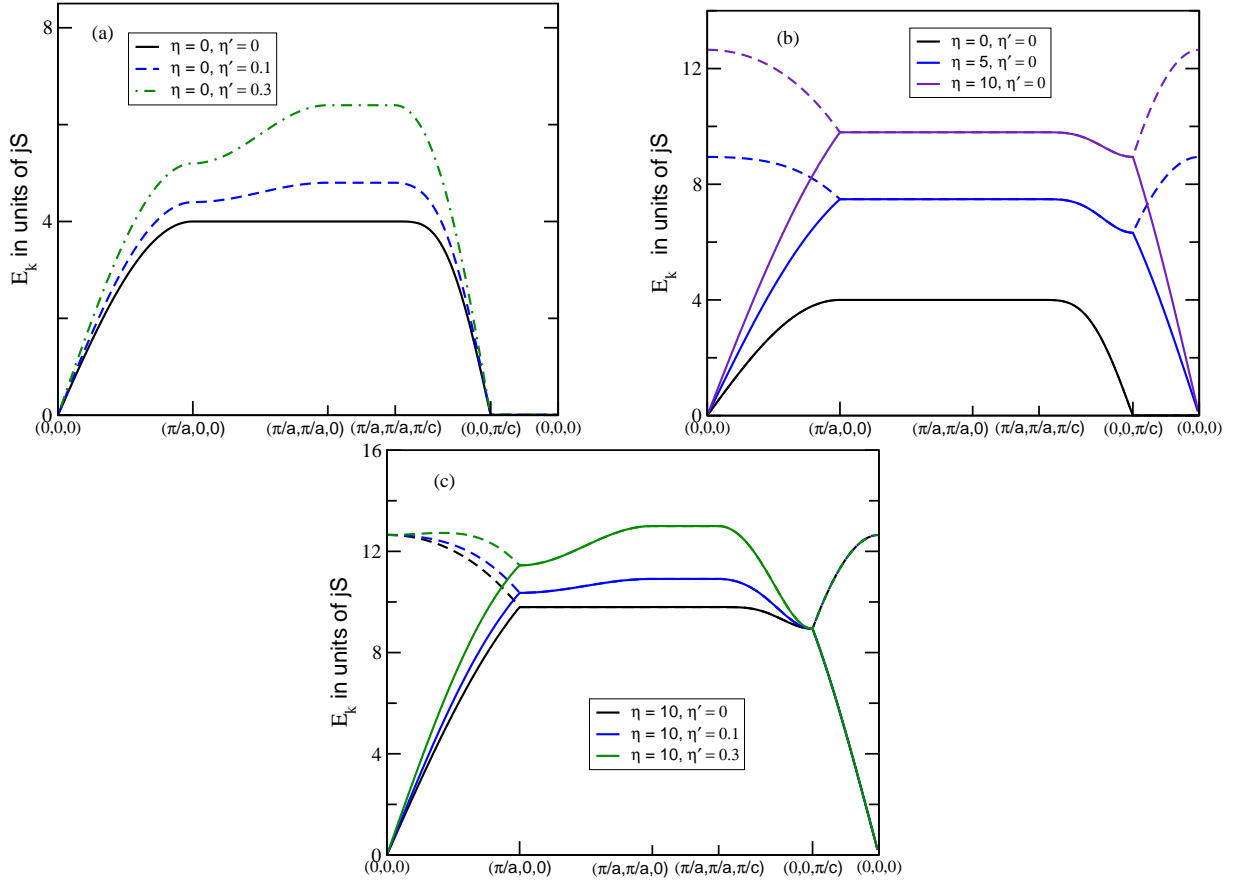

\centering
\includegraphics[width=3.0in,clip]{EnergyAFAFa.eps}
\qquad 
\includegraphics[width=3.0in,clip]{EnergyAFAFb.eps}
\qquad 
\includegraphics[width=3.0in,clip]{EnergyAFAFc.eps}
\caption{\label{fig:dispAFAF} (Color online) Magnon dispersion for acoustic (Goldstone) and optic magnons are shown for 
different values of intra-dimer coupling $\eta=J/j$
and NNN intrabilayer ferromagnetic interaction $\eta^\prime=j^\prime/j$ (a) - (c). Effects of $\eta^\prime$ on the dispersion 
are shown for two different 
values of $\eta$ (a), (c). 
}
\end{figure}
In Fig.~\ref{fig:dispAFAF} we show the magnon dispersions. Since the tetragonal unit cell contains four Cr spins (two up and two down) 
there will be four spin wave (SW) branches, two $\alpha$ and two $\beta$ branches, for each ${\bf k}$. 
In Fig.~\ref{fig:dispAFAF}a we show the results for $\eta=0$, 
when the two bilayers are decoupled. In this limit, branch 1 and branch 2 with frequencies $\omega_{\bf k}^{(1)}$ and  
$\omega_{\bf k}^{(2)}$ correspond 
to two bilayers and are degenerate. For $\eta^\prime=0$, the magnon is dispersionless (within LSWT) from $(\pi/a,0,0)$ to
$(\pi/a, \pi/a,0)$ to $(\pi/a, \pi/a, \pi/c)$. 
The absence of any $k_z$ dependence is obvious as with $\eta=0$ there is no coupling along the $z$-direction. However 
$(\pi/a,0,0)$ to $(\pi/a, \pi/a,0)$ 
independence is peculiar to a NN 2D antiferromagnet. Introduction of magnon-magnon interaction or a nonzero NNN exchange 
coupling $\eta^\prime$ brings in dispersion. For the known case with $J=j^\prime=0$ our system is a 2D antiferromagnet on a square lattice
with only NN interaction $j$.
In that case for the region $(\pi, 0)$ to $(\pi/2,\pi/2)$ 
the dispersion is flat within linear spin wave theory and with $1/S$ corrections. However, 
with $1/S^2$ corrections, magnon energy at $(\pi/a,0)$ is smaller than at $(\pi/2a, \pi/2a)$ (See 
Ref.~\onlinecite{majumdar10, majumdar13} for details).
In our case if we include the 2nd order corrections we will find similar results. However, for our system 
S=3/2, we expect the dip (if there is any) to be smaller. In the limit $J$ and $j$ are zero, $j^\prime$ controls the magnon dispersion. The 
system consists of non-interacting 2D ferromagnetic sheets (square lattice with NN coupling $j^\prime$) with dispersion 
$\omega_{\bf k}=4j^\prime S[1-(\cos (k_x a)+\cos (k_y a))/2]$, which for small ${\bf k}$ takes the well-known result 
$\omega_{\bf k}=2j^\prime S(ka)^2$. 

In Fig.~\ref{fig:dispAFAF}b we see the effect of introducing inter-bilayer 
AF coupling $\eta$ (for 
simplicity we chose $\eta^\prime=0$). Non-zero $\eta$ couples the intra-bilayer modes, leading to acoustic (Goldstone modes, 
$\omega_{\bf k}^{(1)} \rightarrow 0$ as ${\bf k}\rightarrow 0$) 
and optic modes ($\omega_{\bf k}^{(2)} \rightarrow 4\sqrt {\eta}/(4+\eta)$ as ${\bf k}\rightarrow 0$). The new $\alpha$ and $\beta$ modes are linear 
combinations of the old decoupled bilayer $\alpha$ and $\beta$  
modes. The four-fold degenerate modes split into two modes along $(0,0,0)$ to $(\pi/a,0,0)$ and the zero frequency modes along $(0,0,0)$ 
to $(0,0, \pi/c)$ split into acoustic and optic modes.  Interestingly, the modes along $(\pi/a,0,0)$ to $(\pi/a, \pi/a,0)$ to 
$(\pi/a, \pi/a, \pi/c)$ 
are dispersionless  and four-fold degenerate. Finally, in Fig.~\ref{fig:dispAFAF}c, we show how the NNN ferromagnetic coupling 
introduces dispersion 
to these modes, but it does not remove the degeneracy. 

\subsection{Sublattice Magnetization}
\begin{figure}[httb]
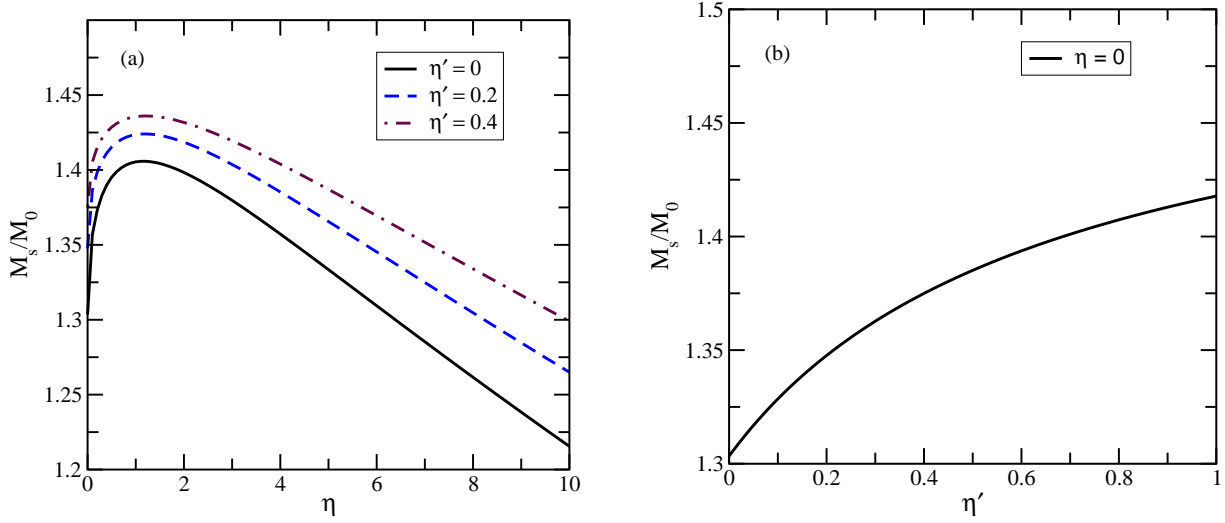

\centering
\includegraphics[width=3.0in,clip]{MagAFAFa.eps}
\qquad 
\includegraphics[width=3.0in,clip]{MagAFAFb.eps}
\caption{\label{fig:magAFAF} (Color online) (a) Normalized sublattice magnetization, $m_s=M_s/M_0$ is shown as a function of the inter-bilayer coupling parameter 
$\eta$ for three different values of NNN interaction $\eta^\prime$. 
(b) The effect of ferromagnetic interaction $\eta^\prime$ on $m_s$ for $\eta=0$ is shown. For large $\eta^\prime$, $m_s$
approaches the classical value 1.5.}
\end{figure}

The effect of QSF in quantum anti-ferromagnets in reducing the mean-field 
(or classical) value of the sublattice magnetization is well known.~\cite{anderson} This effect 
is strongest for small $S$ values and low dimensions. In our system, if we turn off 
the inter-bilayer AF exchange ($\eta=0$), the system reduces to 2D quantum $S=3/2$ system. 
In Fig.~\ref{fig:magAFAF}, 
we discuss the effect of QSF on the scaled magnetization $m_s=M_s/M_0$ where $M_0=g\mu_B$. In Fig.~\ref{fig:magAFAF}a, we see that for 
the decoupled bilayers ($\eta=0$), QSF reduces $m_s$ from 1.5 (classical value) to 1.303 when $\eta=\eta^\prime=0$, a 13\% reduction 
from the classical value. 
When one increases the strength of ferromagnetic NNN coupling the QSF effect is suppressed and the value 
increases towards its classical mean-field value. In Fig.~\ref{fig:magAFAF}b we show the effect of increasing $\eta^\prime$
on the magnetization. Going back to Fig.~\ref{fig:magAFAF}a we find that for $\eta^\prime=0$, magnetization starts at 1.303 
for $\eta = 0$ and increases to 1.406 at 
$\eta=1.2$ and then decreases monotonically. The 2D QSF are suppressed as one introduces inter-bilayer coupling, 
even if it is antiferromagnetic; this is a 3D effect. But when $\eta$ increases further local dimer-related QSFs 
start to dominate and for large values of $\eta$ ($>\sim 6$), the magnetization is smaller than the 2D value.~\cite{majumdar12,majumdar10} 
In general, we find that the effect of nonzero $\eta^\prime$ is to suppress QSF effect on the magnetization.

\subsection{Two-Magnon Density of States (DOS)}

\begin{figure}[httb]
\centering
\includegraphics[width=\textwidth,clip]{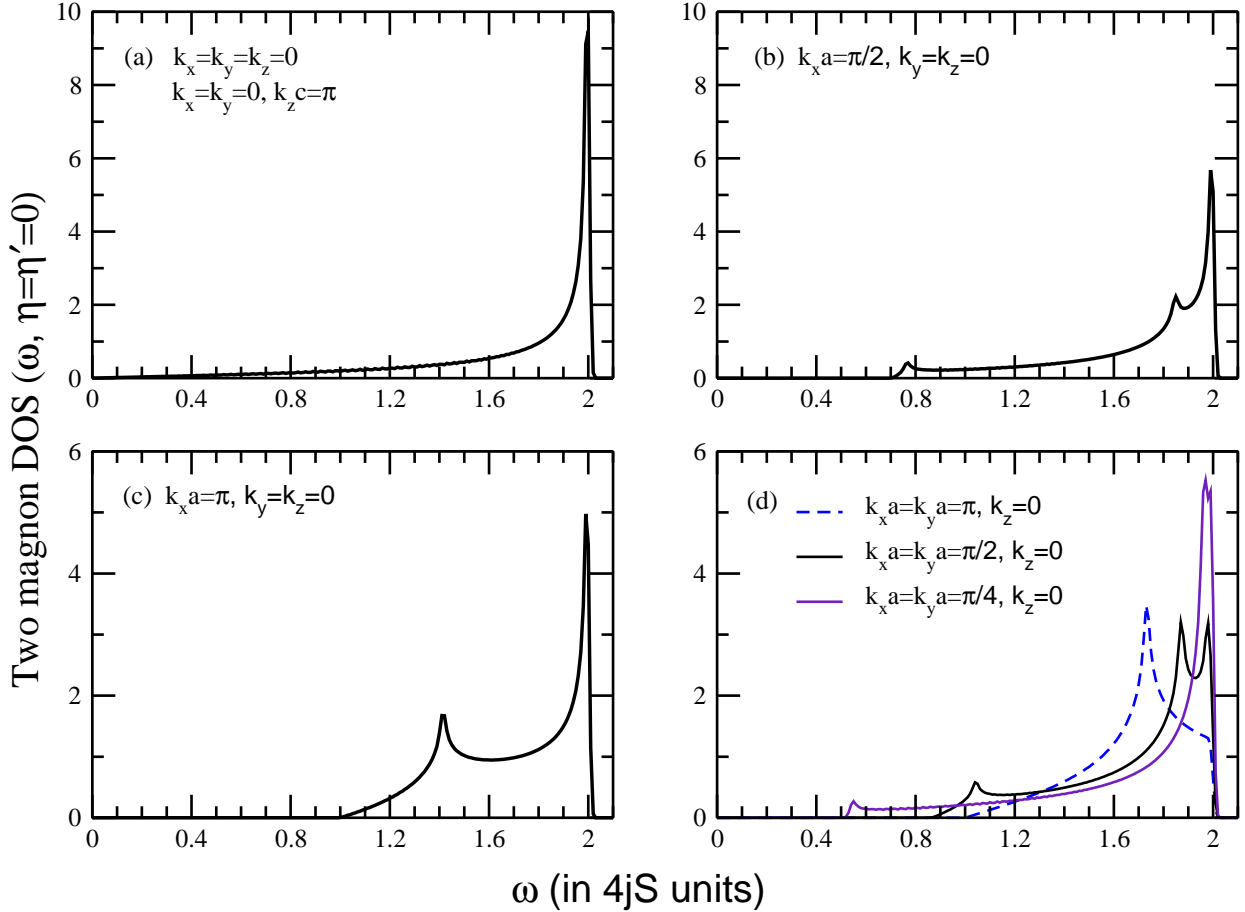}
\caption{\label{fig:AFAFDOS}(Color online) Two-magnon density of states (DOS) for different values of ${\bf k}$ is plotted 
for $\eta=\eta^\prime=0$. In this 
case DOS for the $\alpha$ and $\beta$ magnon branches are identical. The single peak structure at $8jS$ in (a) at the $\Gamma$-point (${\bf k}=0$)
develops two and three peak structures as one goes away from the $\Gamma$-point as seen in (b)-(d).}
\end{figure}

As we have discussed in Sec.~\ref{LSSF} the longitudinal spin-spin correlation function $L_s({\bf k},\omega)$, which is directly probed in 
inelastic scattering measurements depends sensitively on the two-magnon DOS [see Eqs.~\eqref{spin-corr} and \eqref{dos}]. The latter were 
calculated for different ${\bf k}$-values by numerically evaluating Eqs.~\eqref{dos1} - \eqref{dos4}. The sum over the internal 
three-dimensional momenta ${\bf p}$ is done on mesh grid of size $L \times L \times L$, where $L = 256$. A Gaussian function 
of width 0.075 (in units of energy) was used to broaden the $\delta$-function. As seen in Eq.~\eqref{dos}, there are contributions from 
the two branches 1 and 2 in different combinations (11, 22, 21, 12). 
In Fig.~\ref{fig:AFAFDOS}a-d, we present the two-magnon DOS$({\bf k},\omega)$ for $\eta=0$. In the absence of inter-bilayer coupling 
the bilayer modes are degenerate. 
In addition, the $\alpha$ and $\beta$ modes are also degenerate due to symmetry. Thus we have a 4-fold degenerate magnon mode 
[see Fig.~\ref{fig:dispAFAF}a] 
and in this case 
DOS$_{11}({\bf k},\omega)=$DOS$_{22}({\bf k},\omega)=$DOS$_{21} ({\bf k}, \omega)=$DOS$_{12} ({\bf k}, \omega)=$DOS$({\bf k},\omega)/4.$ 
The peak occurs at $8jS$ whereas the peak in the DOS 
of one-magnon excitation is at $4jS$. This factor of 2 is a result of linear spin wave (LSW) approximation. As seen 
in Fig.~\ref{fig:AFAFDOS}, 
DOS$({\bf k}, \omega)$ is independent of $k_z$ as it should be. The single peak structure seen for ${\bf k}=0$ ($\Gamma$-point) 
at $8jS$ develops two and three peak 
structures as one goes away from the $\Gamma$-point. This is seen clearly in Fig.~\ref{fig:AFAFDOS}d. At the Brillouin zone 
boundary $(k_x=k_y=\pi/a)$, the peak appears below the band edge, at $\sim 6.8jS$ and there is a saddle point at $8jS$. 

\begin{figure}[httb]
\centering
\includegraphics[width=\textwidth,clip]{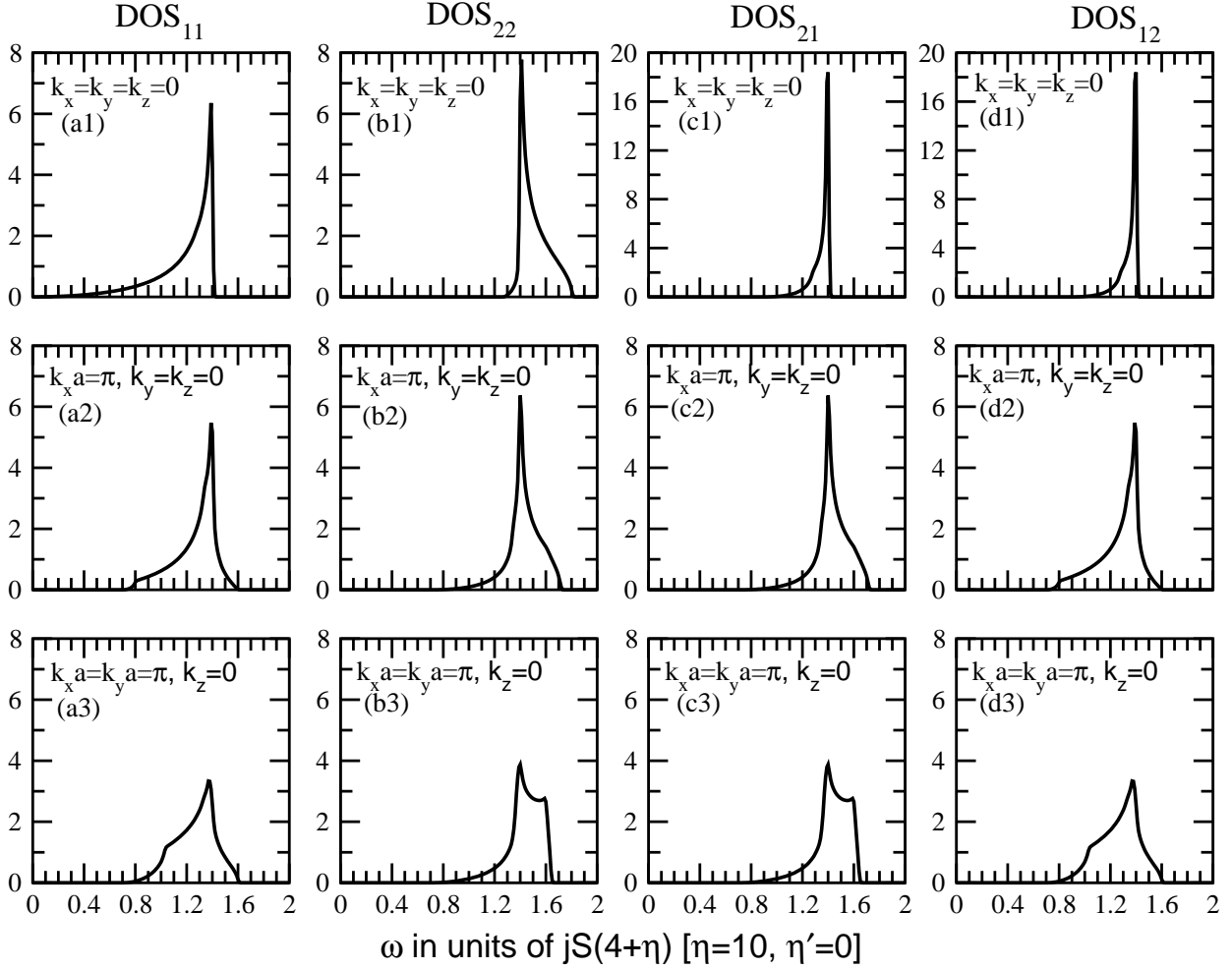}
\caption{\label{fig:AFAFDOS1} Two magnon DOS for different values of $(k_x, k_y)$ with $k_z=0$ is plotted for $\eta=10, \eta^\prime=0$.
(a1) - (d3)
Note that at the $\Gamma$-point (${\bf k}=0$), DOS$_{21}=$DOS$_{12}$ (a1) - (d1)
whereas for $k_x=k_y=\pi/a,k_z=0,$ DOS$_{11}=$DOS$_{12}$ and DOS$_{22}=$DOS$_{21}$ (a3) - (d3).}
\end{figure}

\begin{figure}[httb]
\centering
\includegraphics[width=\textwidth,clip]{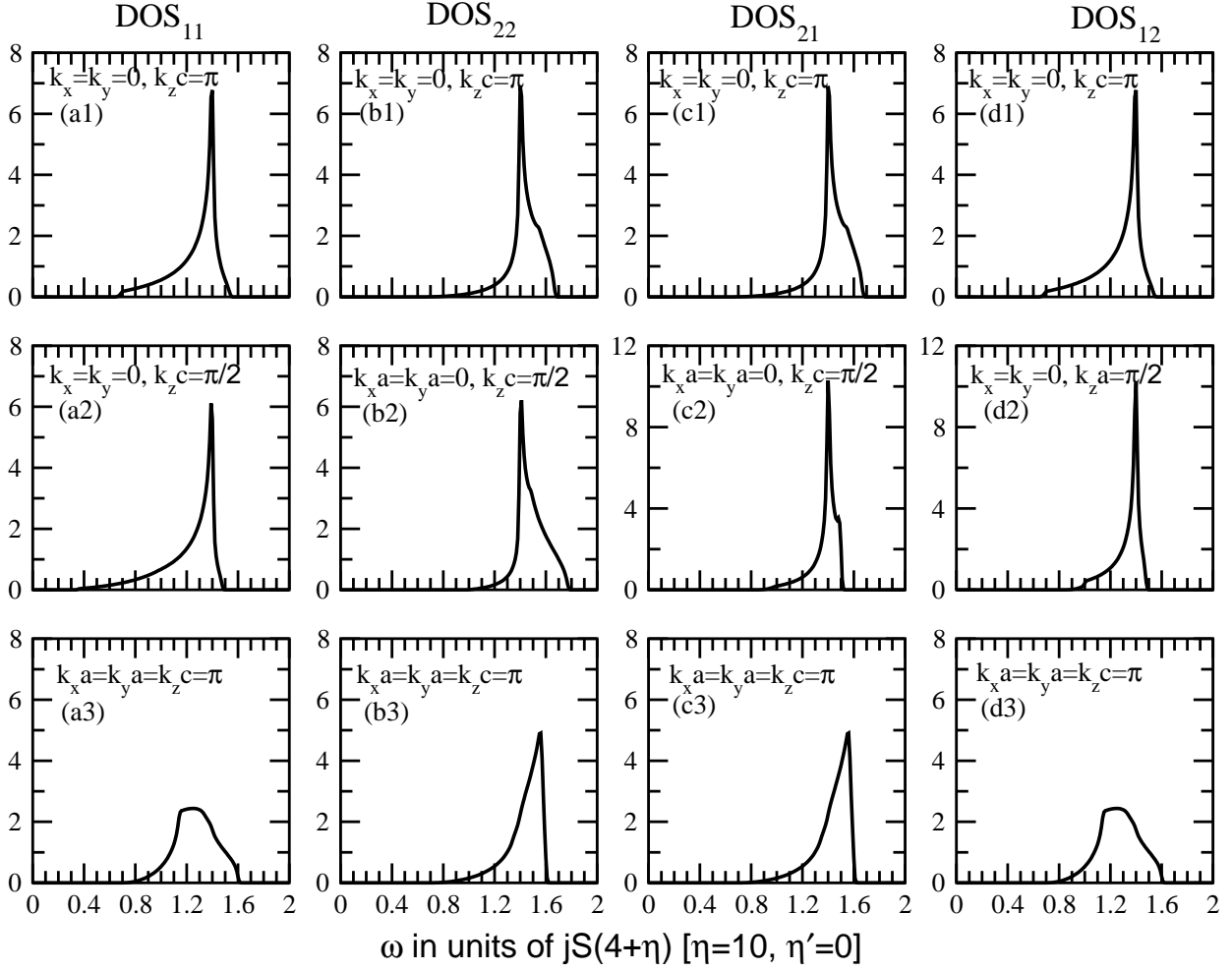}
\caption{\label{fig:AFAFDOS2} $k_z$ dependence on the two magnon DOS is shown in (a1) - (d3) for $\eta=10, \eta^\prime=0$. All the 
peaks occurs
at $19.5 jS$. For $k_x=k_y=0, k_z=\pi/c$ (a1) - (d1) and $k_x a=k_y a=k_z c=\pi$ (a3) - (d3),
DOS$_{11}=$DOS$_{12}$ and DOS$_{22}=$DOS$_{21}$. But for $k_x=k_y=0, k_z=\pi/2c$ (a2) - (d2), all DOS's differ from each other.}
\end{figure}

Next, we discuss the case when inter-bilayer coupling $(J)$ is nonzero. Since in the Cr$_2$XO$_6$ systems, $J$ is much larger 
than the intra-bilayer coupling $(j)$ we choose $\eta=10$ and still keep $\eta^\prime=0$ for simplicity. In 
Fig.~\ref{fig:AFAFDOS1} and Fig.~\ref{fig:AFAFDOS2}, we plot the 
$({\bf k},\omega)$ dependence of DOS$_{11}$, DOS$_{22}$, DOS$_{21}$, and DOS$_{12}$. In Fig.~\ref{fig:AFAFDOS1}, 
we choose $k_z=0$ and study the $(k_x,k_y)$ dependence. 
The dominant feature is a narrow peak at energy $\sim 19.5jS$, whereas the single magnon branches lie between 0 and 
$13jS$ [Fig.~\ref{fig:dispAFAF}c]. The peak comes from the flat part of the single magnon dispersion near $9.75jS$. Since DOS$_{11}$ comes 
only from the acoustic branch it extends below the peak. On the other hand, DOS$_{22}$ comes only from the optic branch, 
it extends above the peak. The mixed contributions DOS$_{12}$ and DOS$_{21}$ have strong ${\bf k}$-dependence. It is interesting to note 
that for $k_x=k_y=k_z=0$ [Fig.~\ref{fig:AFAFDOS1}c1, d1], DOS$_{21}=$DOS$_{12}$. This is because at the $\Gamma$-point, Eqs.~\eqref{dos3} and Eq.~\eqref{dos4} are 
identical. On the other hand, for $k_x=k_y=\pi/a,k_z=0$ [Fig.~\ref{fig:AFAFDOS1}a3-d3], DOS$_{11}=$DOS$_{12}$ and DOS$_{22}=$DOS$_{21}$. 
In Fig.~\ref{fig:AFAFDOS2}, we show the effect of $k_z$ on all four DOS. 
In contrast to the $\eta=0$ case, for $\eta=10$ DOS depends on $k_z$. For example, when $k_z=\pi/c$, all the four DOS differ from 
their corresponding structures when $k_x=k_y=k_z=0$ [Fig.~\ref{fig:AFAFDOS2}a3-d3]. Instead they are identical to the 
spectra when $k_x=\pi/a, k_y=k_z=0$.
Interestingly for $k_x=k_y=0, k_z=\pi/c$ and $k_x a=k_y a=k_z c=\pi$ [Fig.~\ref{fig:AFAFDOS2}a1-d1 and Fig.~\ref{fig:AFAFDOS2}a3-d3],
DOS$_{11}=$DOS$_{12}$ and DOS$_{22}=$DOS$_{21}$. These symmetries are mentioned in Sec.~\ref{TMDOS}.
But for $k_x=k_y=0, k_z=\pi/2c$ [Fig.~\ref{fig:AFAFDOS2}a2-d2] all DOS's are different.

\subsection{Longitudinal Spin-Spin Correlation Function (LSSF)}
\begin{figure}[httb]
\centering
\includegraphics[width=\textwidth,clip]{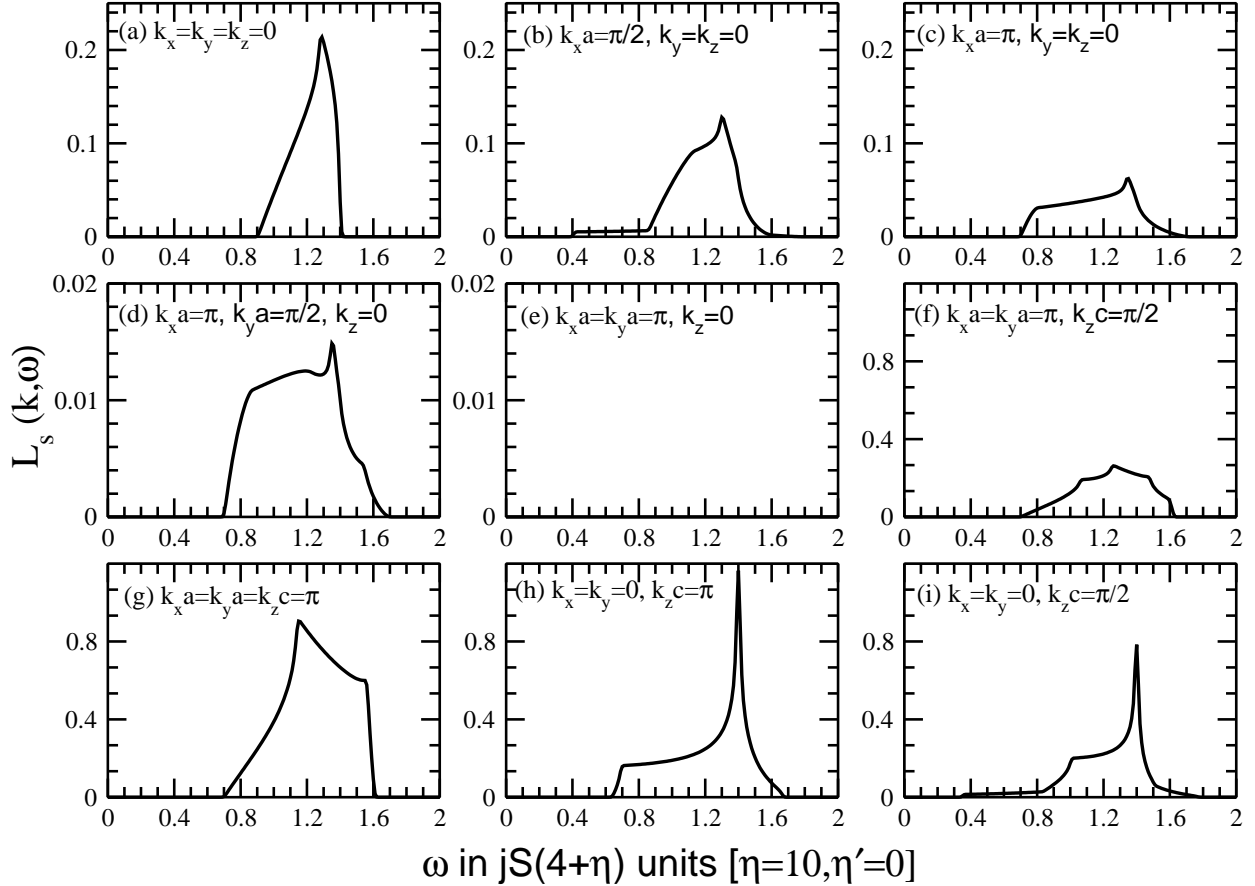}
\caption{\label{fig:SzzAFAF1} Longitudinal spin-spin correlation, ${\cal L}_{s}({\bf k}, \omega)$ for different values of ${\bf k}$ 
is plotted for $\eta=10, \eta^\prime=0$ in (a) - (i). Note that for $k_x=k_y=\pi/a, k_z=0$, ${\cal L}_{s}({\bf k}, \omega)=0$ [Fig.\ref{fig:SzzAFAF1}e]. 
A narrow peak is seen around $19.5jS$.}
\end{figure}

\begin{figure}[httb]
\centering
\includegraphics[width=\textwidth,clip]{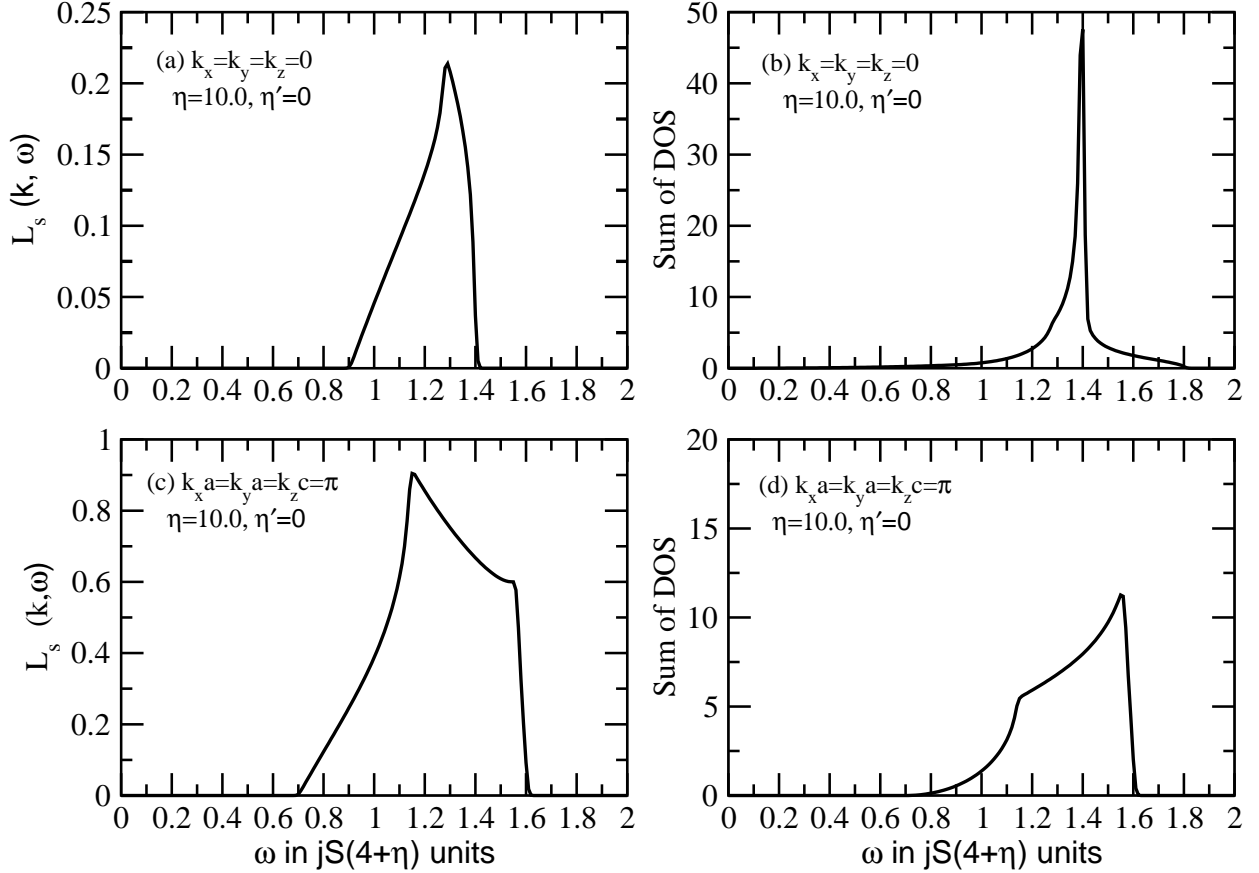}
\caption{\label{fig:SzzDOS} Longitudinal spin-spin correlation ${\cal L}_{s}({\bf k}, \omega)$ and the sum of the density of states of 
four magnon branches are plotted for $\eta=10, \eta^\prime=0$ and two different values of ${\bf k}=0$ (a) - (b) 
and $k_xa=k_ya=k_zc=\pi$ (c) - (d). 
The plots display the effects of form factors in ${\cal L}_{s}({\bf k}, \omega)$.}
\end{figure}

In Fig.~\ref{fig:SzzAFAF1}a-i, we show the ${\bf k}$-dependence of LSSCF ${\cal L}_s({\bf k},\omega)$. As seen in 
Eq.~\eqref{spin-corr}, contributions from different two-magnon 
excitations get weighted by the associated form factors ${\cal D}^{ij}_{{\bf k},{\bf k+p}}$. This leads to different energy dependence of 
LSSCF compared to that of the total two-magnon DOS. For example, as seen in Fig.~\ref{fig:SzzAFAF1}e, for $k_x=k_y=\pi/a, k_z=0$, LSSCF 
vanishes and becomes nonzero as we increase $k_z$. In Fig.~\ref{fig:SzzDOS} we show both ${\cal L}_s({\bf k},\omega)$ and the sum of 
the four DOS$_{ij}({\bf k}, \omega)$ 
for ${\bf k}=0$ and $k_xa=k_ya=k_zc=\pi$. For both the ${\bf k}$ values the effect of the form factors is very significant. 

\begin{figure}[httb]
\centering
\includegraphics[width=\textwidth,clip]{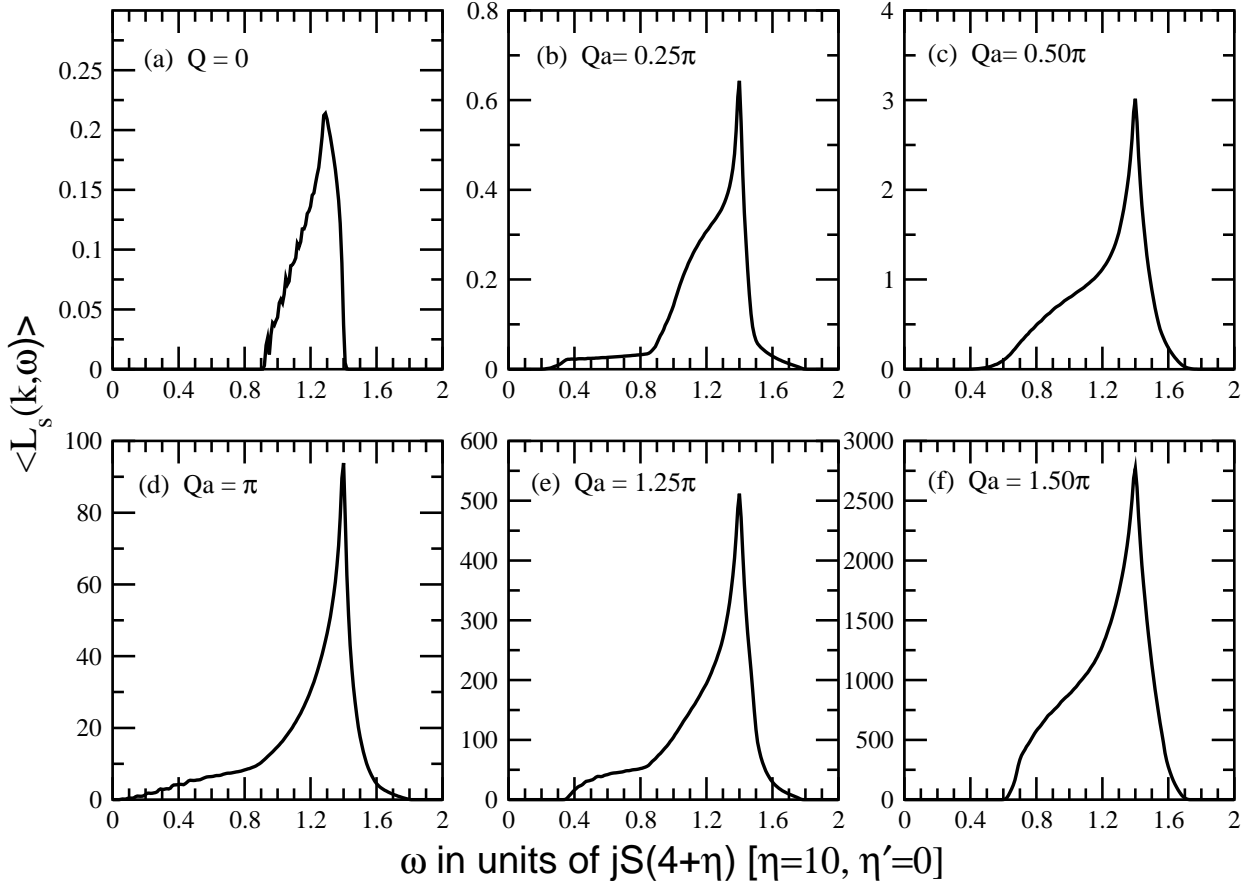}
\caption{\label{fig:PowAvg}Powder-averaged longitudinal spin-spin correlation function for $\eta=10, \eta^\prime=0$ for 
$Q=0, 0.25\pi/a, 0.50\pi/a, \pi/a, 1.25\pi/a, 1.50\pi$/a is shown in (a) - (f). A narrow peak occurs at $19.5jS$ - the  
intensity scale of this peak grows with the magnitude of ${\bf k}$.}
\end{figure}

Finally, we plot 
the angular average of ${\cal L}_s({\bf k},\omega)$ for different magnitudes of ${\bf k}$ in Fig.~\ref{fig:PowAvg}. 
For these plots, Eq.~\eqref{powavg} was numerically 
evaluated by summing over the angles $\theta,\phi$. 
For each $\omega$ about 270 million points were evaluated. This is what is observed in a inelastic neutron scattering experiment
from a powder sample. The generic feature is a narrow peak seen at $19.5jS$ [at $\omega=1.4jS(4+\eta)$] with a small broad peak at 
lower energies. Also note how the intensity scale grows with the magnitude of ${\bf k}$.

\section{\label{sec:conclusions}Conclusions}
In this paper we have discussed the magnon dispersion, two-magnon density of states, and longitudinal spin-spin correlation 
function in the leading order approximation, in systems described by coupled bilayers where both intra ($j$) and inter-bilayer ($J$) 
nearest neighbor (NN) couplings are antiferromagnetic. Although the particular spin system we have studied is Cr$_2$TeO$_6$, 
which contain Cr$^{3+}$ ions with spin-3/2, our formalism is general and can be applied to any spin-$S$. We have also investigated 
how a small intra-bilayer NNN ferromagnetic coupling ($j^\prime$) affects the above properties.

One of the interesting features of our calculation is the non-monotonic $\eta=J/j$ dependence of the reduced sublattice magnetization
$m_s=M_s/M_0$. In the classical limit (mean-field) $m_s =1.5$ for $S=3/2$. When $\eta = 0$, that is in the decoupled bilayer limit, 
quantum spin fluctuations (QSF) reduce $m_s$ to 1.303 but as we increase $\eta$, $m_s$ first increases and equals to 1.406 when 
$\eta \sim 1$ and then decreases and becomes smaller than the decoupled bilayer value when $\eta > 6$. For Cr$_2$TeO$_6$, $\eta$ is 
estimated to be $\sim 10$ (interacting quantum spin-dimer limit) and $m_s$ differs substantially from its classical value. 
The presence of nonzero $j^\prime$, QSF effects are suppressed.

Due to the quasi 2D geometry and local (NN) inter bilayer AF coupling, magnon dispersion shows a flat region over a large 
part of the 2D Brillouin zone. This results in a sharp peak in the one-magnon density of states (DOS) near $\sim 10 jS$ for $\eta=10$,
not at the band maximum which occurs at $\sim 12 jS$, but closer. The two-magnon DOS also shows sharp peaked structure for most 
of the values of the total momentum ${\bf k}$. The two-magnon peak appears $\sim 1.4(4+\eta)jS=19.5jS$. The longitudinal 
spin-spin correlation function, ${\cal L}_s({\bf k},\omega)$ function depends both on the two-magnon spectrum and the 
Bogoliubov amplitudes and phases. In fact, for certain ${\bf k}$, ${\cal L}_s({\bf k},\omega)$ vanishes even if the two-magnon 
DOS does not. Experiments in single crystal samples should test the results of this theoretical predictions. Unfortunately, 
large single crystal samples of Cr$_2$TeO$_6$ are not available, most of the neutron experiments are done in powder samples. 
In this case, one measures the energy dependence of ${\cal L}_s({\bf k},\omega)$ averaged over the angular components of ${\bf k}$, 
that is for different magnitudes of  $|{\bf k}|(=Q)$. Again one finds a peak structure for most of the values of $Q$, again at $~19.5jS$, 
but the main effect of increasing $Q$ is seen in the increase of the total intensity. These predictions can be checked experimentally.

In a subsequent paper, we will discuss the case when intra-bilayer exchange is ferromagnetic (e.g. in Cr$_2$WO$_6$  and Cr$_2$MoO$_6$) 
and compare this with the present case. Here the QSF effects are absent in the limit $\eta = 0$, but start to increase as
one increases $\eta$.
\section{Acknowledgment}
We acknowledge the use of HPC cluster at GVSU, supported by the National Science Foundation 
Grant No. CNS-1228291 that have contributed to the research results reported within this paper. SDM would like 
to thank Dr. Xianglin Ke for stimulating discussions.

\appendix
\section{\label{SScorr} Total spin $S_z$ in terms of $\alpha$ and $\beta$ magnons}
\bea
S_z({\bf k})&=& -\frac 1{2}\frac 1{\sqrt{4N}}\sum_{{\bf p,q}}\delta ({\bf k}+{\bf p}-{\bf q})\times \non \\
&\Big[&\{[f_{1{\bf k}}+f_{4{\bf k}}]C_{1\bf p}C_{1\bf q}
-[f_{2{\bf k}}+f_{3{\bf k}}]\zeta_{1\bf p}^*\zeta_{1\bf q}S_{1\bf p} S_{1\bf q}\}
\alpha_{\bf p}^{(1)\dagger}\alpha_{\bf q}^{(1)} \non \\
&+& \{[f_{1{\bf k}}+f_{4{\bf k}}]C_{2\bf p}C_{2\bf q}
-[f_{2{\bf k}}+f_{3{\bf k}}]\zeta_{2\bf p}^*\zeta_{2\bf q}S_{2\bf p} S_{2\bf q}\}
\alpha_{\bf p}^{(2)\dagger}\alpha_{\bf q}^{(2)}\non \\
&+& \{[f_{1{\bf k}}+f_{4{\bf k}}]S_{1\bf p}S_{1\bf q}
-[f_{2{\bf k}}+f_{3{\bf k}}]\zeta_{1\bf p}\zeta_{1\bf q}^*C_{1\bf p} C_{1\bf q}\}
\beta_{-\bf q}^{(1)\dagger}\beta_{-\bf p}^{(1)}\non \\
&+& \{[f_{1{\bf k}}+f_{4{\bf k}}]S_{2\bf p}S_{2\bf q}
-[f_{2{\bf k}}+f_{3{\bf k}}]\zeta_{2\bf p}\zeta_{2\bf q}^*C_{2\bf p} C_{2\bf q}\} 
\beta_{-\bf q}^{(2)\dagger}\beta_{-\bf p}^{(2)} \non \\
&-&\{[f_{1{\bf k}}+f_{4{\bf k}}]C_{1\bf q}S_{1\bf p}
-[f_{2{\bf k}}+f_{3{\bf k}}]\zeta_{1\bf p}^*\zeta_{1\bf q}C_{1\bf p} S_{1\bf q}\}
\alpha_{\bf q}^{(1)}\beta_{-\bf p}^{(1)} \non \\
&-& \{[f_{1{\bf k}}+f_{4{\bf k}}]C_{1\bf p}S_{1\bf q}
-[f_{2{\bf k}}+f_{3{\bf k}}]\zeta_{1\bf p}^*\zeta_{1\bf q}C_{1\bf q} S_{1\bf p}\}
\alpha_{\bf p}^{(1)\dagger}\beta_{-\bf q}^{(1)\dagger}\non \\
&-&\{[f_{1{\bf k}}+f_{4{\bf k}}]C_{2\bf q}S_{2\bf p}
-[f_{2{\bf k}}+f_{3{\bf k}}]\zeta_{2\bf p}^*\zeta_{2\bf q}C_{2\bf p} S_{2\bf q}\}
\alpha_{\bf q}^{(2)}\beta_{-\bf p}^{(2)} \non \\
&-& \{[f_{1{\bf k}}+f_{4{\bf k}}]C_{2\bf p}S_{2\bf q}
-[f_{2{\bf k}}+f_{3{\bf k}}]\zeta_{2\bf p}^*\zeta_{2\bf q}C_{2\bf q} S_{2\bf p}\}
\alpha_{\bf p}^{(2)\dagger}\beta_{-\bf q}^{(2)\dagger} \non \\
&+&\{[f_{1{\bf k}}-f_{4{\bf k}}]C_{1\bf p}C_{2\bf q}
+[f_{2{\bf k}}-f_{3{\bf k}}]\zeta_{1\bf p}^*\zeta_{2\bf q}S_{1\bf p} S_{2\bf q}\}
\alpha_{\bf p}^{(1)\dagger}\alpha_{\bf q}^{(2)} \non \\
&+& \{[f_{1{\bf k}}-f_{4{\bf k}}]C_{2\bf p}C_{1\bf q}
+[f_{2{\bf k}}-f_{3{\bf k}}]\zeta_{2\bf p}^*\zeta_{1\bf q}S_{2\bf p} S_{1\bf q}\}
\alpha_{\bf p}^{(2)\dagger}\alpha_{\bf q}^{(1)}\non \\
&+& \{[f_{1{\bf k}}-f_{4{\bf k}}]S_{1\bf q}S_{2\bf p}
+[f_{2{\bf k}}-f_{3{\bf k}}]\zeta_{1\bf q}\zeta_{2\bf p}^*C_{1\bf q} C_{2\bf p}\}
\beta_{-\bf q}^{(1)\dagger}\beta_{-\bf p}^{(2)}\non \\
&+& \{[f_{1{\bf k}}-f_{4{\bf k}}]S_{1\bf p}S_{2\bf q}
+[f_{2{\bf k}}-f_{3{\bf k}}]\zeta_{1\bf p}^*\zeta_{2\bf q}C_{1\bf p} C_{2\bf q}\} 
\beta_{-\bf q}^{(2)\dagger}\beta_{-\bf p}^{(1)} \non \\
&-&\{[f_{1{\bf k}}-f_{4{\bf k}}]C_{1\bf q}S_{2\bf p}
+[f_{2{\bf k}}-f_{3{\bf k}}]\zeta_{2\bf p}^*\zeta_{1\bf q}C_{2\bf p} S_{1\bf q}\}
\alpha_{\bf q}^{(1)}\beta_{-\bf p}^{(2)} \non \\
&-& \{[f_{1{\bf k}}-f_{4{\bf k}}]C_{1\bf p}S_{2\bf q}
+[f_{2{\bf k}}-f_{3{\bf k}}]\zeta_{1\bf p}^*\zeta_{2\bf q}C_{2\bf q} S_{1\bf p}\}
\alpha_{\bf p}^{(1)\dagger}\beta_{-\bf q}^{(2)\dagger}\non \\
&-&\{[f_{1{\bf k}}-f_{4{\bf k}}]C_{2\bf q}S_{1\bf p}
+[f_{2{\bf k}}-f_{3{\bf k}}]\zeta_{1\bf p}^*\zeta_{2\bf q}C_{1\bf p} S_{2\bf q}\}
\alpha_{\bf q}^{(2)}\beta_{-\bf p}^{(1)} \non \\
&-& \{[f_{1{\bf k}}-f_{4{\bf k}}]C_{2\bf p}S_{1\bf q}
+[f_{2{\bf k}}-f_{3{\bf k}}]\zeta_{2\bf p}^*\zeta_{1\bf q}C_{1\bf q} S_{2\bf p}\}
\alpha_{\bf p}^{(2)\dagger}\beta_{-\bf q}^{(1)\dagger}
\Big].
\eea
\bibliography{CrMW}

\end{document}